# Effects of biaxial strain on the electronic structures and band topologies of group-V elemental monolayers

Jinghua Liang, Long Cheng, Jie Zhang, Huijun Liu[*]

*Key Laboratory of Artificial Micro- and Nano-Structures of Ministry of Education and School of Physics and Technology, Wuhan University, Wuhan 430072, China*

Using first-principles calculations, we systematically investigate the electronic structures and band topologies of four kinds of group-V elemental (P, As, Sb and Bi) monolayers with buckled honeycomb structure. It is found that all these monolayers can change from semiconducting to semimetallic under compressive strain. If a tensile strain is however applied, the P, As and Sb monolayers undergo phase transition from topologically trivial to non-trivial regime, whereas the topological insulating nature of Bi monolayer remains unchanged. With tunability of the band gaps and band topologies, it can be expected that these elemental monolayers could be promising candidates for future optoelectronic and spintronic applications.

In recent years, many efforts have been devoted to the study of a new class of quantum materials known as topological insulators (TIs), which is also called quantum spin Hall (QSH) phase in two-dimensional (2D) systems. Such materials have bulk band gaps and topologically protected boundary states which offer exciting possibility in the applications of spintronics [1], quantum computation [2, 3] and catalysis [4, 5]. The TIs are characterized by a so-called $\mathbb{Z}_2$ topological invariant $v$, which equals to 1 for TIs but becomes 0 for trivial insulators [6, 7]. It has been generally accepted that the TI nature is driven by strong spin-orbital coupling (SOC), which plays an important role in determining the band order around the Fermi level. So far, only the HgTe/CdTe [8, 9] and InAs/GaSb [10] quantum wells are well-established as QSH insulators experimentally. Much effort has therefore been devoted to the search of more new QSH systems. On the theoretical side, it was previously shown that group-IV elemental monolayers, including graphene [11], silicene, germanene, [12] and stanene [13], are all QSH insulators. Further studies showed that their band gaps can be tuned by external strain [14], chemical functionalization [14, 15], and substrate [16].

---

[*] Author to whom correspondence should be addressed. Electronic mail: phlhj@whu.edu.cn

Group-V elemental monolayers have attracted a lot of interests as new 2D materials, especially after the fabrication of phosphorene (α-P) exfoliated from black phosphorous [17]. Recently, Zhang *et al.* theoretically [18] predicted the existence of two novel group-V elemental monolayers, namely, the arsenene (As monolayer) and the antimonene (Sb monolayer). Moreover, they also reported the successful fabrication of antimonene using van der Waals epitaxy growth [19]. It is worthy to note that group-V monolayers could have interesting topological properties. For example, Liu *et al.* [20] predicted that phosphorene (α-P) can be converted into TI by applying electric field. It was also shown that the band topology of Sb nano-films can be tuned by strain, electric field [21], and film thickness [22]. In the case of Bi, the ultrathin (111) films were found to be characterized by a non-trivial $\mathbb{Z}_2$ independent of the film thickness [23, 24]. Indeed, the non-trivial topological nature [25, 26] of ultrathin Bi films grown on substrate has been revealed experimentally. Moreover, the band topology of Bi thin film is found to be robust against strain and electrical field [27]. It was further predicted that, by chemical modification, Bi/Sb halide and hydride monolayers are both stable QSH insulators with extraordinarily large bulk gaps ranging from 0.32 eV to a record value of 1.08 eV [28]. Although the electronic structures of some group-V elemental monolayers have been reported, the band topologies of the whole family remains less studied, especially for their response to the strain. In this work, using first-principles calculations, we give a comprehensive study of the effects of biaxial strain on the electronic structures and band topologies of four kinds of group-V elemental (P, As, Sb, and Bi) monolayers. We will consider both compressive and tensile strain in a large range from −20% to 20%.

Our first-principles calculations have been performed within the framework of density functional theory (DFT) [29], as implemented in the Vienna *ab initio* simulation package (VASP) [30]. The all-electron projector augmented wave (PAW) method [31, 32, 33] is used to treat the ion-electron interactions. The energy cutoff for plane wave expansion is 400 eV, and a $\Gamma$-centered ***k***-point mesh of 12×12×1 is used for the Brillouin zone integration. All the monolayers are modeled by unit cells repeated periodically along the in-plane directions, and a vacuum distance of 20 Å is adopted in the out-of-plane direction to avoid the interactions between the monolayers and their periodic images. The atomic positions are fully relaxed until the magnitude of the force acting on each atom is less than 0.01 eV/Å. The SOC is explicitly included in our calculations. As the standard DFT usually underestimates the band gap, we adopt the Heyd-Scuseria-Ernzerhof (HSE) screened Coulombic hybrid density functional [34] to calculate the band structures and evaluate the $\mathbb{Z}_2$ topological invariant.

Elemental bulk materials of group-V elements can have many allotropes. At ambient conditions, the most stable states of As, Sb and Bi elements exhibit a layered structure with space group of $R\bar{3}m$, as shown in Figure 1(a). Such structure can be also found in bulk P, but at high pressure [35, 36]. As illustrated in Figure 1(b), one can mechanically exfoliate elemental monolayer from the bulk material, which is very similar to obtaining graphene from graphite. Moreover, we see that the group-V monolayers also exhibit honeycomb structure with two atoms in the unit cell, but are buckled for stabilization [18, 37, 38, 39]. To discuss the structural properties, we define the buckling angle $\theta$ by $\cos\theta = h/d$, where $d$ and $h$ are the bond length and buckling distance (Figure 1(c)). Compared with the low-buckled honeycomb structures of group-IV elemental monolayers with $\theta \approx 90°$ [11, 12, 13], we find that all the group-V elemental monolayers show high-buckled structures with $\theta \approx 55°$. In Figure 2, we show the bond length $d$ and the buckling distance $h$ as a function of the biaxial strain, which is defined as $\varepsilon = (a-a_0)/a_0$. Here $a = \sqrt{3}d\sin\theta$ and $a_0$ is the equilibrium lattice constant. At equilibrium, we see that both the bond length and the buckling distance of these elemental monolayers increase with increasing atomic mass, that is, $d_P > d_{As} > d_{Sb} > d_{Bi}$ and $h_P > h_{As} > h_{Sb} > h_{Bi}$. When the tensile strain is applied ($\varepsilon > 0$), we see that the bond lengths $d$ of these monolayers increase monotonically while the buckling distances $h$ decrease with increasing strain. The reason is that the stretch of in-plane lattice constant will result in out-plane relaxation of atoms. On the other hand, when the tensile strain is applied ($\varepsilon < 0$), the buckling distances $h$ increase monotonically with increasing strain, while there is a local minimum of the bond length $d$ at $\varepsilon \approx -10\%$. Such an observation suggests possible phase transition at larger compressive strain, and we will come back to this point later.

It is well known that inducing strain is an effective way in band engineering. Our extensive calculations on the four kinds of group-IV elemental monolayers indicate that, except for the Bi monolayer, the other three systems show quite similar behaviors as a function of strain. Here we take the As monolayer as an example (results for P and Sb monolayers are given in Figure S1 and S2 of the Supplementary Material). Figure 3 shows the energy band structures of As monolayer at several typical biaxial strains. To have a better comparison, the results with and without SOC are both shown. At equilibrium, we see that the As monolayer is an indirect gap semiconductor if SOC is not considered (up panel of Figure 3(b)). The valence band maximum (VBM) of As monolayer appears at the $\Gamma$ point, while the conduction band minimum (CBM) locates at about half of the $\Gamma M$ line. Such observation is different from that found for the low-buckled group-IV elemental monolayers [11, 12, 13], where both the VBM and CBM appear at the $K$ point. Detailed analysis of the band projections indicates that the valence and conduction bands of As monolayer near the Fermi level are mostly composed of the $p$ orbitals. Unlike that of group-IV elemental monolayers [40], the high-buckled structure of As monolayer causes the $p_{xy}$ orbitals hybrid with the the $p_z$ orbitals. If the SOC is explicitly included in the calculations, we see the degeneracies of the $p_{xy}$-type valence and conduction bands at the $\Gamma$ point are lifted. Moreover, we see that the $p_z$-type valence and conduction bands are lower than the corresponding $p_{xy}$-type bands. When a compressive strain is applied to the As monolayer, we see from Figure 3(a) that both the $p_z$-type valence and conduction bands move away from the Fermi level. Furthermore, we find that As monolayer is converted to a semimetal when the value of compressive strain is larger than 10%. If instead a tensile strain is applied, there is an increase of the buckling angle $\theta$, and the $p_z$-type valence and conduction bands both move towards the Fermi level. Moreover, the CBM tends to move to the $\Gamma$ point, making the As monolayer a direct gap semiconductor. At a tensile strain of 15%, the CBM coincides with the VBM and the gap is closed (Figure 3(c)). It is interesting to find that the two $p_z$-type bands at the $\Gamma$ point are inverted if the tensile strain is further increased. Indeed, we see from Figure 3(d) that when $\varepsilon = 20\%$, the two $p_z$-type bands cross each other at about 1/4 $\Gamma K$ away from the $\Gamma$ point, leading to six Dirac cone-like band structures in the first Brillouin zone (BZ). Moreover, a small gap is opened at the Dirac cone when the SOC is considered, as indicated in the lower panel of Fig. 3(d).

The observed band inversion of As monolayer is very similar to those of TIs [41], which makes it reasonable to expect that the As monolayer could convert to a QSH insulator when the applied tensile strain is larger than a critical value. This can be verified by directly calculating the $\mathbb{Z}_2$ invariants $v$. Following the approach proposed by Fu and Kane [42], for a 2D system with space inversion symmetry such as our investigated monolayers, the $\mathbb{Z}_2$ invariants $v$ can be calculated from the parities of wave function at four time-reversal-invariant $k$ points ($k_i$), which are the $\Gamma$ point and three equivalent $M$ points in the hexagonal BZ of group-V monolayers. That is, $(-1)^v = \prod_i^4 \delta(k_i) = \delta(\Gamma)\delta(M)^3$ with $\delta(k_i) = \prod_{n=1}^N \xi_{2n}^i$. Here $\xi = \pm$ stands for even (+) or odd (−) parity eigenvalues, and $N$ is half the number of occupied bands. Indeed, our calculation confirms that the As monolayer is a trivial insulator with $v = 0$ when the applied tensile strain is less than 15%, while becomes a QSH insulator with $v = 1$ when $\varepsilon > 15\%$.

In Figure 4, we summarize the calculated band gaps of four kinds of group-V monolayers as a function of biaxial strain. In the case of P, As, and Sb, we see that all of them exhibit semiconductor-to-semimetal transition under compressive strain and indirect-to-direct gap transition under small tensile strain. Such tunability suggests that they could be promising materials for applications in optoelectronics. Furthermore, by directly calculating the $\mathbb{Z}_2$ invariants, we confirm that P, As and Sb monolayers can realize the QSH states under sufficient large tensile strain, which suggests that they may have potential application in spintronic devices. With the increasing atomic mass and SOC strength, we see that the critical strain for the topological transition is 19%, 15%, and 13.5% for the P, As and Sb monolayers, respectively. For the Bi monolayer, however, the much stronger SOC strength can drastically change the band structure (see Figure S3 of the Supplementary Material). Note that previous calculations [27] adopting Perdew-Burke-Ernzerhof functional predict that the band gap of Bi monolayer remains open within a strain range from −6% to 6%. This is however a bit different from our results. As can be seen from Figure 4 and Figure S3, the semiconducting Bi monolayer becomes semimetallic when the compressive strain is larger than 8%. On the other hand, the band gap closes at a tensile strain of 5%, and opens again when the tensile strain is further increased. Based on the parity analysis, we find that the Bi monolayer remains topological non-trivial at smaller (< 8%) compressive strain, or if tensile strain is applied (even though its band gap is closed at $\varepsilon = 5\%$). In fact, it can be seen from Figure 5 that when the tensile strain is increased from 1% to 2%, the first and second highest valence bands are inverted around the $\Gamma$ point, rendering the parities of the VBM and CBM are both "+". As a consequence, the $\mathbb{Z}_2$ invariant of Bi monolayer remains unchanged although the VBM and CBM are inverted when the tensile strain is larger than 5%.

In summary, our first-principles calculations demonstrate that all the four kinds of group-V elemental monolayers (P, As, Sb, and Bi) can convert from semiconducting to semimetallic when a compressive strain larger than 10% is applied. Interestingly, we observe a topological phase transition for the P, As, and Sb monolayers at decreasing critical tensile strain of 19%, 15%, and 13.5%, respectively. For the Bi monolayer with much stronger SOC strength, however, the band topology remains non-trivial at compressive strain less than 8% or if tensile strain is applied. In experiments, the interfacial strain is unavoidable in thin films grown by epitaxial method supported on substrates [19, 25, 26] because of lattice mismatch and charge transfer between the film and substrate. Alternatively, compressive or tensile strain can be realized through thin film shrinkage [43] or by stretching films with atomic force microscopy (AFM) manipulation [44]. Our theoretical work thus offers a feasible route to effectively control the electronic structures and band topologies of group-V elemental monolayers.

**Acknowledgements**

We thank financial support from the National Natural Science Foundation (grant No. 11574236) and the "973 Program" of China (Grant No. 2013CB632502).

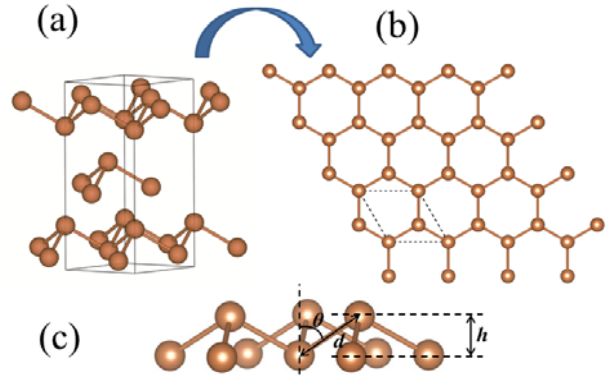

**Figure 1** (a) Crystal structure of bulk P, As, Sb and Bi with space group $R\bar{3}m$. (b) Top-view of the group-V elemental monolayers with honeycomb lattice. The dashed lines indicate the unit cell. (c) Side-view of the group-V elemental monolayers. The bond length $d$, the buckling distance $h$, and the buckling angle $\theta$ are indicated.

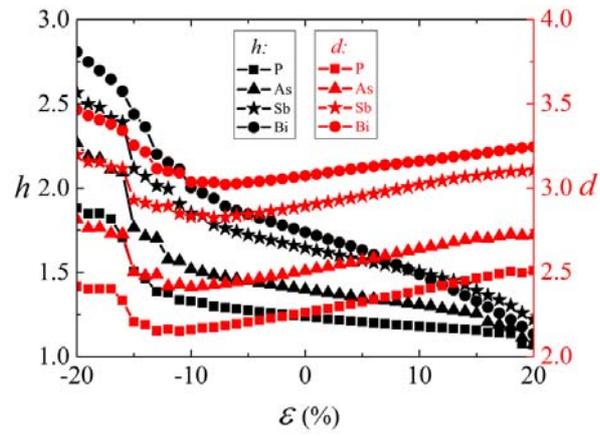

**Figure 2** The bond length $d$ (red) and buckling distance $h$ (black) of P, As, Sb and Bi monolayers as a function of biaxial strain $\varepsilon$.

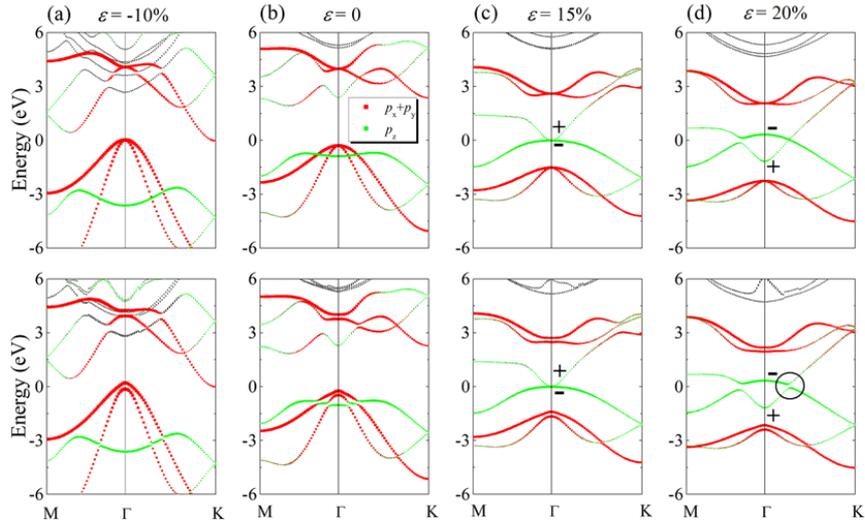

**Figure 3** The orbital-decomposed band structures of As monolayer at different biaxial strain. The upper and the lower panels indicate calculations without and with SOC, respectively. The red and green squares represent the weights of $p_{xy}$ and $p_z$ character, respectively. The Fermi level is at 0 eV. The symbols "+" and "−" indicate the parities of $p_z$-type bands at the $\Gamma$ point.

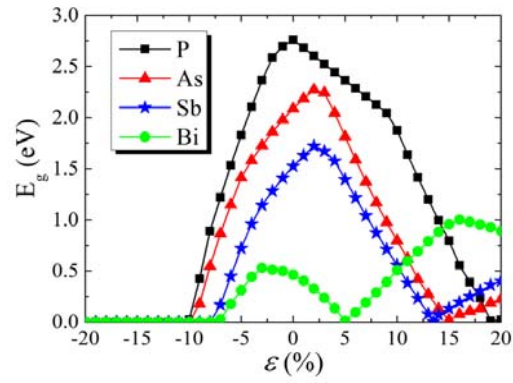

**Figure 4** The calculated band gaps of P, As, Sb and Bi monolayers as a function of biaxial strain $\varepsilon$.

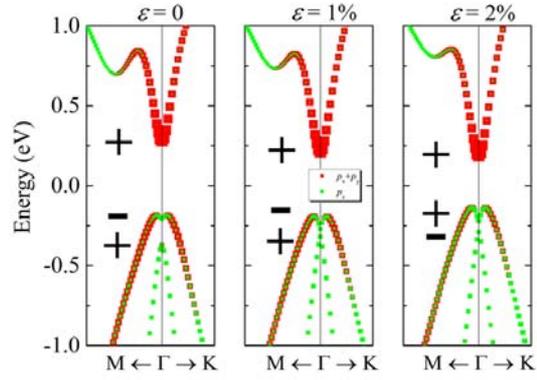

**Figure 5** The orbital-decomposed band structures of Bi monolayer, compared with those at two tensile strains of 1% and 2%. The red and green squares represent the weights of $p_{xy}$ and $p_z$ character. The Fermi level is at 0 eV. The parities of the lowest conduction band, the first and the second highest valence bands at the Γ point are labeled.